\documentclass[preprints,article,accept,pdftex,moreauthors]{Definitions/mdpi} 
\firstpage{1} 
\pubvolume{1}
\issuenum{1}
\articlenumber{0}
\pubyear{2025}
\copyrightyear{2025}
\datereceived{ } 
\daterevised{ } 
\dateaccepted{ } 
\datepublished{ } 
\hreflink{https://doi.org/} 

\Title{Visualization of high-intensity laser-matter interactions in virtual reality and web browser}

\TitleCitation{Visualization of high-intensity laser-matter interactions in virtual reality and web browser}

\Author{Martin Matys $^{1,}$*\orcidA{}, James P. Thistlewood $^{1,2}$\orcidB{}, Mariana Kecová $^{1}$\orcidC{}, Petr Valenta $^{1}$\orcidD{}, Martina~Greplová Žáková $^{1}$\orcidE{}, Martin Jirka $^{1,3}$\orcidF{}, Prokopis Hadjisolomou $^{1}$\orcidG{}, Alžběta~Špádová~$^{1,3}$\orcidH{}, Marcel~Lamač $^{1}$\orcidI{}  and Sergei ~V. ~Bulanov $^{1}$\orcidJ{}}

\AuthorNames{Martin Matys, James P. Thistlewood, Mariana Kecová, Petr Valenta, Martina Greplová Žáková, Martin Jirka, Prokopis Hadjisolomou, Alžběta Špádová, Marcel Lamač and Sergei V. Bulanov}

\AuthorCitation{Matys, M.; Thistlewood, J.P.; Kecová, M.;  M.; Valenta, P.; Greplová Žáková, M.; Jirka, M.; Hadjisolomou, P.; Špádová, A.; Lamač, M.; Bulanov, S.V.}

\address{%
$^{1}$ \quad ELI Beamlines Facility, The Extreme Light infrastructure ERIC, Za Radnici 835, 25241 Dolni Brezany,
Czech Republic.\\
$^{2}$ \quad Department of Physics, University of Oxford, OX1 3PU, United Kingdom.\\
$^{3}$ \quad Faculty of Nuclear Sciences and Physical Engineering, Czech Technical University in Prague, Brehova 7,	11519 Prague, Czech Republic.}
\corres{Correspondence: Martin.Matys@eli-beams.eu}

\abstract{We present the Virtual Beamline (VBL) application, an interactive web-based platform for visualizing high-intensity laser-matter interactions using particle-in-cell (PIC) simulations, with future potential for experimental data visualization. These interactions include ion acceleration, electron acceleration, $\gamma$-flash generation, electron-positron pair production, and attosecond and spiral pulse generation.  Developed at the ELI Beamlines facility, VBL integrates a custom-built WebGL engine with WebXR-based Virtual Reality (VR) support, allowing users to explore complex plasma dynamics in non-VR mode on a computer screen or in fully immersive VR mode using a head-mounted display. The application runs directly in a standard web browser, ensuring broad accessibility. VBL enhances the visualization of PIC simulations by efficiently processing and rendering four main data types: point particles, 1D lines,  2D textures, and 3D volumes. By utilizing interactive 3D visualization, it overcomes the limitations of traditional 2D representations, offering enhanced spatial understanding and real-time manipulation of visualization parameters such as time steps, data layers, colormaps. Users can interactively explore the visualized data by moving their body or using a controller for navigation, zooming, and rotation. These interactive capabilities improve data exploration and interpretation, making VBL a valuable tool for both scientific analysis and educational outreach. The visualizations are hosted online and freely accessible on our server, providing researchers, the general public, and broader audiences with an interactive tool to explore complex plasma physics simulations. By offering an intuitive and dynamic approach to large-scale datasets, VBL enhances both scientific research and knowledge dissemination in high-intensity laser-matter physics.}


\begin{document}



\section{Introduction}

  The rapid advancement of high-power laser technology has led to the development of numerous multi-petawatt laser facilities worldwide \cite{Danson2019}. This progress has fueled growing interest in compact, laser-driven accelerators for charged particles \cite{Tajima1979, Esarey2009, Bulanov2002, Daido2012, Macchi2013, Bulanov2014, Passoni2019} and radiation sources \cite{Teubner2009, Krausz2009}. Recent breakthroughs in laser electron acceleration have demonstrated energy gains reaching 10 gigaelectronvolts (GeV) over just tens of centimeters \cite{Gonsalves2019, Aniculaesei2023}, while ion acceleration experiments have achieved proton beams with energies up to 150 MeV \cite{Ziegler2024}. These advancements have enabled a wide range of applications, including inertial nuclear fusion \cite{Roth2001,Atzeni2002}, medical applications \cite{Bulanov2014}, and security scanning and probing \cite{Romagnani2005,Albert2016}.  High-power lasers are also instrumental in generating high-brightness photon sources \cite{Bulanov2013} and high-order harmonic generation through laser interactions with gases \cite{Pirozhkov2017} and solid-density targets \cite{Teubner2009, Lamaifmmodecheckcelsevcfi2023}. These advancements open new frontiers in extreme field physics, enabling studies of superstrong electromagnetic fields in quantum regimes \cite{Mourou2006, Marklund2006, DiPiazza2012, Gonoskov2022}. Notable achievements include the experimental creation of electron-positron pairs from vacuum \cite{Burke1997}. Furthermore, theoretical predictions suggest that multi-photon Compton scattering can efficiently generate $\gamma$-ray flashes \cite{Ridgers2012, Nakamura2012, Lezhnin2018,Vyskocil2020, Hadjisolomou2023}, with initial experimental confirmation already reported \cite{Pirozhkov2024}.
The continuous development of laser applications is closely linked to advancements in laser pulse engineering. Optimizing pulse characteristics, such as steepening the pulse front \cite{Vshivkov1998, Matys2022}, increasing peak intensity \cite{Jirka2021_nas}, minimizing prepulses \cite{Dover2023}, and reducing pulse duration from femtoseconds to attoseconds and beyond \cite{Li2025}, is crucial for pushing the boundaries of laser-matter interaction.

As these laser-driven processes become increasingly complex, analyzing the vast and multidimensional datasets generated by simulations and experiments requires advanced visualization techniques. Traditional methods, such as static figures and videos, are inherently limited in how much data they can represent, often failing to capture the full spatial and temporal dynamics of high-dimensional simulations. To overcome these limitations, interactive 3D environments, including Virtual Reality (VR), provide an intuitive way to explore plasma dynamics, particle trajectories, plasma structures, and field structures in real time. These environments can be extended to 4D or multi-D by incorporating temporal evolution \cite{KIM2017} and additional dimension variable (e.g. energy \cite{Igarashi2025}), offering even deeper insight into complex physical phenomena. 


Scientific visualizations of laser-matter interactions and plasma physics using VR are increasingly featured at conferences \cite{Danielova2019,Trindade2024}. The integration of VR enhances these visualizations by enabling dynamic user interaction with complex datasets, moving beyond static depictions. As adoption grows, VR-based visualization is expected to become a standard tool for scientific analysis, presentation, and collaborative research. By allowing users to navigate simulations through movement, zooming, rotation, and time control, VR and 4D visualization provide an intuitive way to explore and understand complex datasets.

To support this approach, we have developed the Virtual Beamline (VBL) application at ELI Beamlines facility \cite{Danielova2019}, providing an interactive and accessible platform for scientific visualization. The latest version integrates a custom-built WebGL \cite{webGL} application with VR support powered by WebXR \cite{webXR}. This design allows VBL to be hosted on a server and accessed by users over the internet, where it runs via a standard web browser. When VR mode is enabled, the application runs in a Head-Mounted Display (HMD), providing an immersive experience. For users without a VR HMD, it can also be accessed on a regular 2D computer screen.

\begin{figure}[H]
 	\centering
 	\isPreprints{}{
 		\begin{adjustwidth}{-\extralength}{0cm}
 			} 
 		\includegraphics[width=1\linewidth]{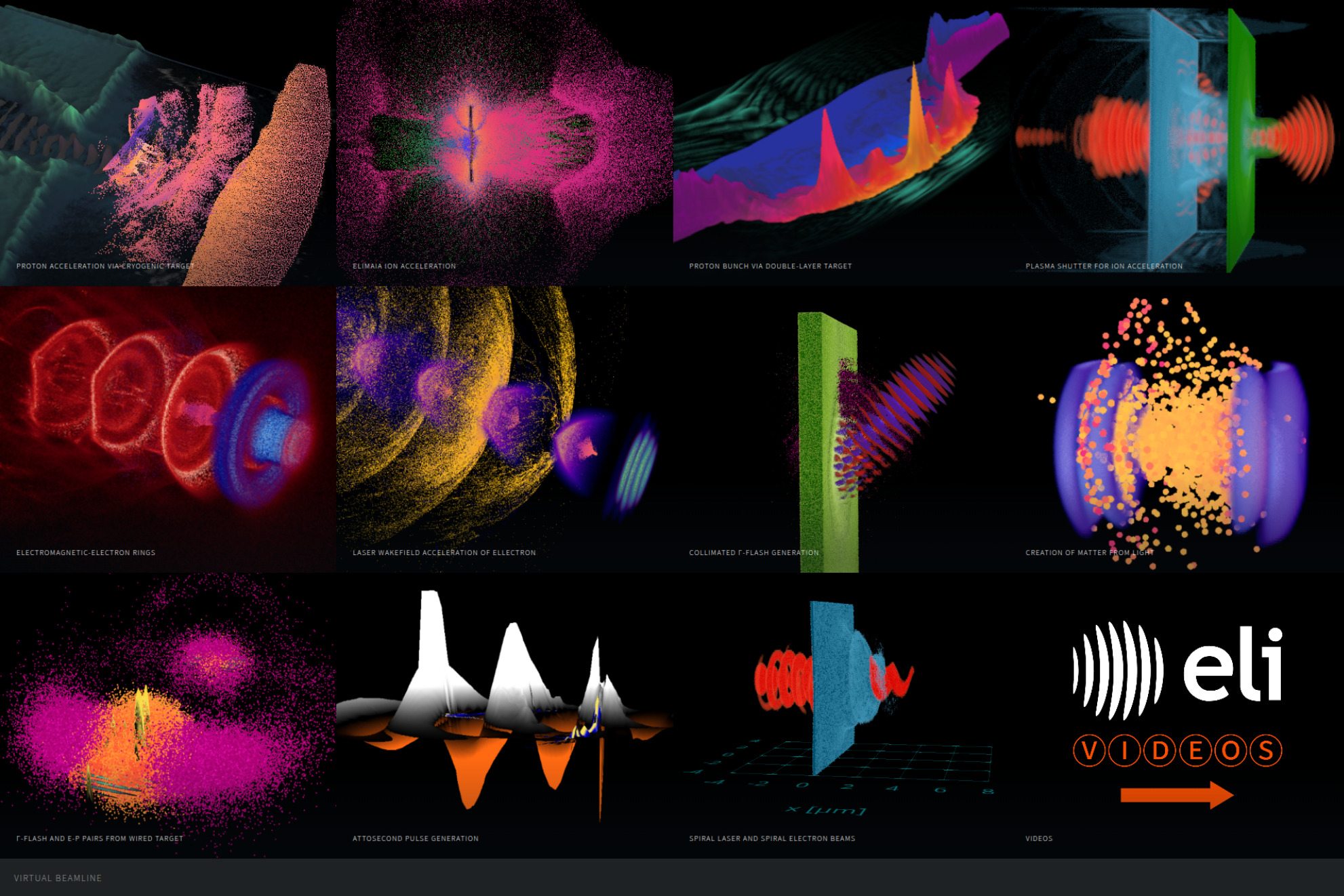}
 		\isPreprints{}{
 		\end{adjustwidth}
 		} 
 	\caption{The current main page of the VBL site (online at \cite{vbl_home_page}) with the available visualizations.}
 	\label{fig1}
 \end{figure}

Both experimental research and theoretical studies using computer simulations generate large volumes of data \cite{Kelling2025} that must be processed, analyzed, and effectively communicated. Rather than storing all data in raw form, it is more practical to preserve only selected portions as interactive visualizations, which are easier to interpret while remaining processable by computers for further analysis. To facilitate this, our repository of visualizations is publicly available \cite{vbl_home_page} (see Fig. \ref{fig1}), currently featuring 11 interactive visualizations based on data simulated at the ELI Beamlines facility.

 Our primary focus is on visualizing data from particle-in-cell (PIC) simulations of high-intensity laser-matter interactions. However, the framework is flexible and can support the visualization of various types of mesh and point data.
 
The paper is organized as follows. The VBL application and its specifications are described in Section \ref{sec2}, with subsections detailing various aspects, including data representation, visualization workflow, user interaction, future development and related VR approaches, covered in Sections \ref{sec21}, \ref{sec22}, \ref{sec23}, \ref{sec24} and \ref{sec25}, respectively. Section \ref{sec3} presents visualizations of various topics in high-intensity laser-matter interactions, categorized into four groups, each described in a separate section, with a brief introduction outlining their applications. Section \ref{sec31} covers ion acceleration, Section \ref{sec32} focuses on electron acceleration, Section \ref{sec33} discusses $\gamma$-flash generation and electron-positron pair production, and Section \ref{sec34} explores attosecond and spiral pulse generation, including the formation of spiral electron beams. The final discussion is presented in Section \ref{sec4}. Appendix \ref{appendix} provides a detailed description of the controls for the VBL application.


\section{Methods} \label{sec2}
The VBL application at ELI Beamlines facility utilizes our web-based interactive 3D visualization framework to render simulation datasets (and experimental data in the future). The application runs in a standard web browser and includes support for VR viewing, offering users a novel and immersive way to view the data. Built on an in-house rendering engine making use of the WebGL standard \cite{webGL}, the framework not only renders datasets but also offers simultaneous view-ports, textual and numerical information, and a GUI featuring timeline animation controls and layer visibility management. Additionally, it integrates graphical elements powered by D3.js \cite{d3js} for plotting animated graphs and legends.

The previous version of our VBL application \cite{Danielova2019} used WebVR \cite{webVR}, which enabled applications to communicate with VR hardware and deliver immersive experiences in a web browser. However, as the experimental WebVR API was deprecated in 2022, it has been largely replaced by its successor, WebXR \cite{webXR}. Consequently, we have rewritten the application's code to use WebXR, which is significantly more widely supported than WebVR was. 

\subsection{Data Representation}\label{sec21}
To effectively visualize simulation data, the system processes and renders four main data types: point particles, 1D lines, 2D textures, and 3D volumes. The data are represented by their spatial positions (XYZ coordinates for particles and grid-based structures for textures and volumes) and by their color, which is mapped to a specific parameter, typically energy or momentum for particles and plasma density or amplitude of electromagnetic fields for textures and volumes. Additionally, for particles, another parameter can be encoded into the alpha channel and used for particle size scale. If particles retain consistent identifiers across time frames, their trajectories can be visualized as 1D lines, providing insight into their motion over time.

A single visualization can include up to four distinct 3D volumetric layers, which can be displayed simultaneously or toggled on and off individually. Different colormaps can be prepared for the volumetric data and adjusted interactively during the visualization runtime.

When a 2D dataset is used, the third (vertical) dimension can represent the particle/texture parameter (which may be the same or different from the parameter mapped to color), creating a pseudo-3D effect. An example of using two different parameters is shown in Section \ref{sec313}, where mean energy in a grid cell is represented by color, while density is mapped to vertical height.

\subsection{Visualization workflow}\label{sec22}
To streamline rendering and ensure consistency across visualizations, our workflow standardizes data transformation, storage, and rendering techniques. Our scripts for data transformation into binary buffers take input data in .h5 format, which is the native format for the PIC codes Smilei \cite{DEROUILLAT2018} or Osiris \cite{OSIRIS}. This format is structurally similar to .sdf, which is used by another PIC code, EPOCH \cite{Arber2015}. Consequently, the preprocessing step often begins by converting user data into .h5 files with a standardized structure. The data is then separated into individual files for different time steps. 

To optimize performance, large datasets can be reduced by cropping regions that will not be visualized, skipping grid points, or interpolating data into a less dense grid. Since PIC simulations typically employ high-density grids to maintain numerical precision, the extra resolution is often unnecessary for visualization purposes. Similarly, the number of visualized particles can be reduced either through random filtering, where only a certain percentage of the original data is retained, or selective filtering, such as filtering based on energy. 

Before transformation, the input data can be previewed in external software to allow users to better understand its multidimensional structure and identify the most relevant attributes for visualization. The data is then processed by our data transformation tool, and the resulting binary buffers are stored on a web server, acting as the data source for the visualization engine. Additionally, advanced shading and rendering techniques like ambient occlusion \cite{occlusion}, 
which enhances spatial depth by simulating how ambient light diminishes in shaded areas, can be applied on demand to improve realism.

\begin{figure}[H]
	\centering
	\isPreprints{}{
		\begin{adjustwidth}{-\extralength}{0cm}
			} 
		\includegraphics[width=1\linewidth]{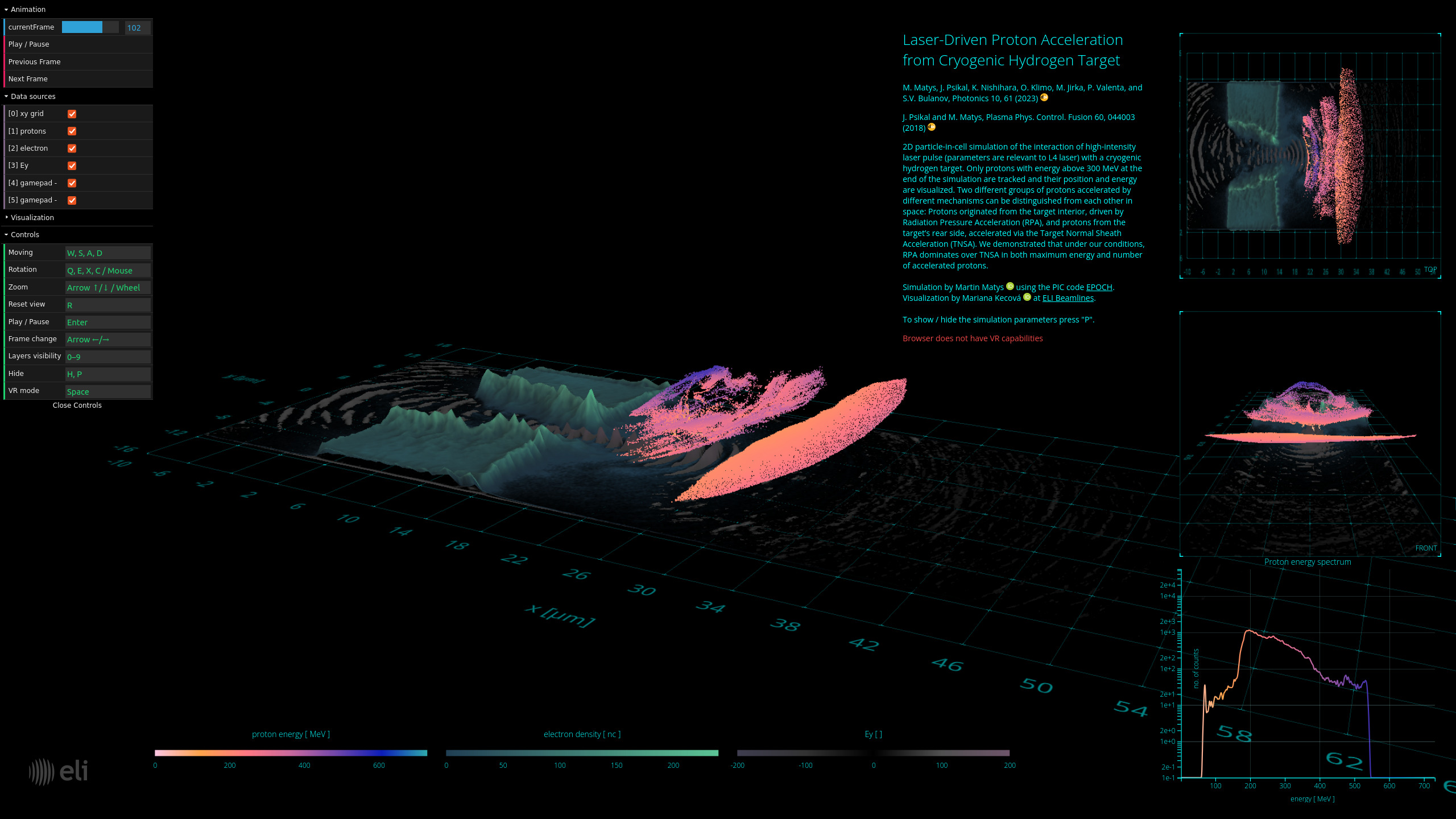}
		\isPreprints{}{
		\end{adjustwidth}
		} 
	\caption{Example of a visualization outside of VR mode (from section \ref{sec311}) displaying multiple viewpoints and graphical elements like graphs and legends.}
	\label{fig2}
\end{figure}

With the data processed, the next step is to configure a scene description for visualization. The data is linked to the visualized layers, and coordinate grids can be added to fit the data. In VR mode, a single viewpoint is used, following the position and direction of the attached HMD. Outside of VR mode, multiple viewpoints can be displayed on the screen, typically a primary interactive viewpoint, shown as the full-background layer, along with smaller windows presenting static viewpoints from specific perspectives (e.g., front or top views). Additionally, various graphical and text elements can be incorporated, such as a visualization title and description, colormaps, legends, and graphs. See Fig. \ref{fig2} for an example. The system utilizes D3.js \cite{d3js} for rendering graphs and legends and dat.GUI \cite{dat.gui} for the user interface, enabling animation controls and layer visibility management.

Lastly, the completed application is uploaded to a web server for online access or a data server is emulated locally. Once the visualization is rendered and deployed, users can interact with the data through both traditional desktop controls and immersive VR environments using an HMD.

\subsection{User Interaction in VR and Non-VR Modes}\label{sec23}

Outside of VR mode, the application is operated using a mouse and keyboard on a computer screen. Users can navigate within the visualization, zoom in and out, and rotate the view. An interactive GUI panel provides additional controls, allowing users to pause the visualization, adjust the time frame, and toggle individual visualization layers on or off. This panel also influences VR mode, enabling partial external control over the visualization displayed in the HMD. This functionality is particularly useful during guided excursions, where an operator can adjust settings in real-time while explaining the visualization to a user inside the HMD. For instance, the operator can toggle layers or select a specific time frame to highlight key aspects of the visualization.

\begin{figure}[H]
	\centering
	\isPreprints{}{
		\begin{adjustwidth}{-\extralength}{0cm}
			} 
		\includegraphics[width=1\linewidth]{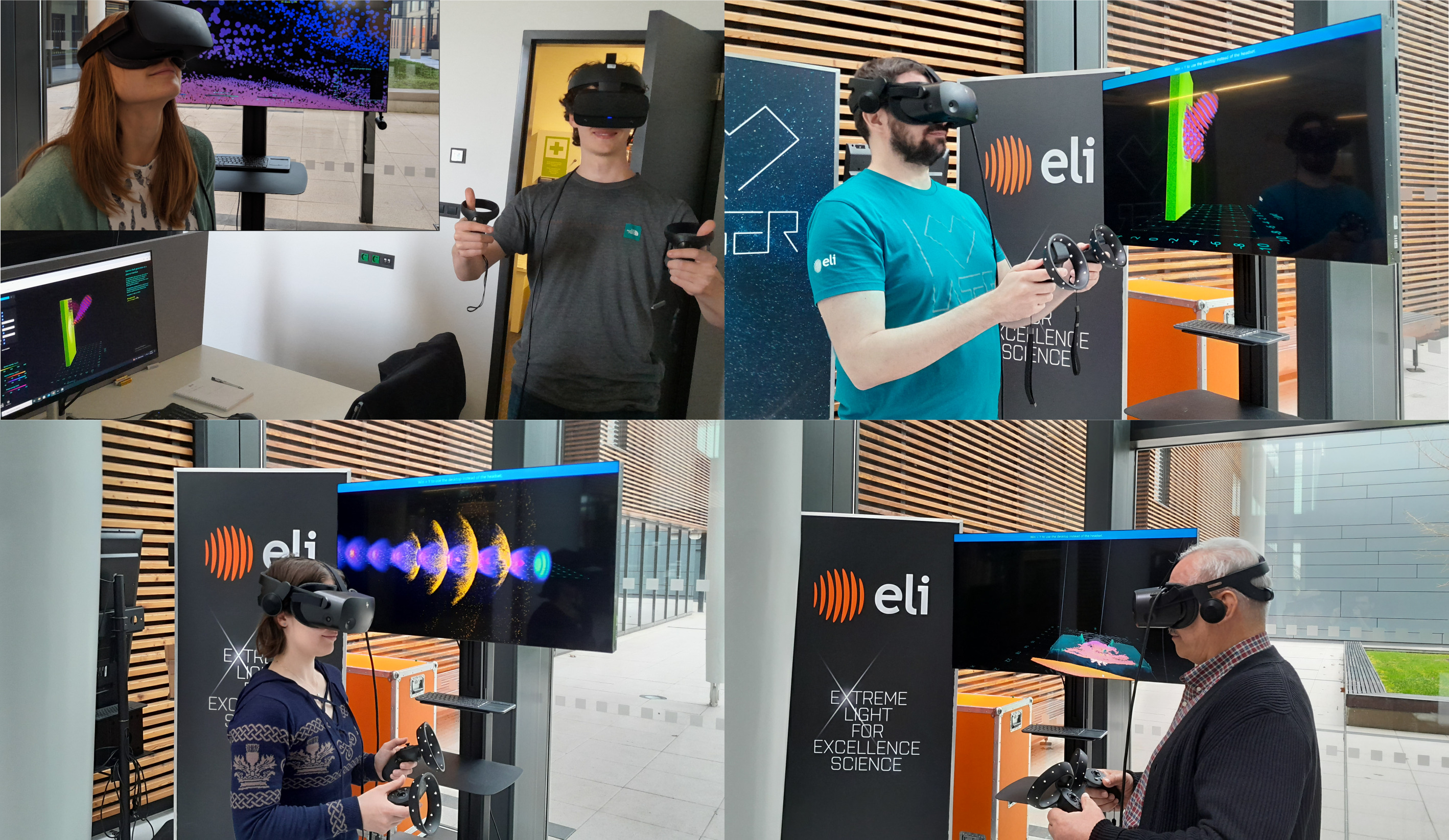}
		\isPreprints{}{
		\end{adjustwidth}
		} 
	\caption{Some of the authors of this paper utilizing VR stations at ELI Beamlines facility.}
	\label{fig_VR_station}
\end{figure}

Inside VR mode, the user’s HMD and controllers are continuously tracked, allowing the viewpoint to respond dynamically. Users can move naturally within the visualization by adjusting their head and body position, creating an immersive and seamless interaction (see users operating the VR stations in Fig. \ref{fig_VR_station}). The VR controllers provide additional functionality, enabling navigation, zooming, and time control. Recently, two new features were introduced to enhance user engagement: horizontal rotation around the center and time manipulation during playback, enabling users to fast-forward or rewind the visualization. These features provide greater control and interactivity, making the VR experience more immersive and dynamic.

The detailed descriptions of the keyboard and VR controller controls are in the appendix \ref{appendix}.

\subsection{Future development}\label{sec24}
Currently, we rely on manual effort to identify relevant data for visualization from the vast datasets generated by simulations and experiments. A trained individual preprocesses the data and uses several Jupyter notebooks to convert it into binary buffers, set up the visualization scene, and configure elements such as color schemes and animation parameters to produce the final application. This process allows us to create new visualizations from local data at ELI ERIC, or from external sources upon agreement. To streamline this workflow, further automation will be necessary in the future, potentially extending to a fully automated process guided by the user. Additionally, feedback from the processed data could be used to optimize subsequent simulations and experiments.

Looking ahead, the growing complexity and volume of data in plasma research point to a future role for artificial intelligence (AI) in data interpretation and adaptive feedback. The foundational neural network model \cite{Hopfield1982} demonstrated how distributed systems can perform associative memory and pattern recognition tasks. This idea was extended in Chapter 11 of Ref. \cite{Tajima1989}, which envisioned neural computation for predicting and controlling plasma behavior. Early applications in space plasma contexts used neural networks to forecast solar wind velocities \cite{Hernandez1993}. More recently, AI and machine learning have been applied to laser–plasma accelerators and control systems, showing promise for experimental optimization and data-driven analysis \cite{Shalloo2020, Jalas2021, Karniadakis2021, Dopp2023, Feister2023, Goodman2023, Loughran2023}. These developments emphasize the potential value of integrating AI approaches into VBL environments in the future.
  \subsection{Related VR approaches}\label{sec25}
Various approaches have been applied to visualize (laser-)plasma simulation data using VR worldwide. One such system, CompleXcope, has been used to visualize both simulation and experimental data from the Large Helical Device for fusion research in Japan \cite{OHTANI_2011,Ohtani2016}. Originally based on the CAVE system \cite{Cruz1993}, CompleXcope projects stereo images onto three walls and the floor, surrounding the user, who wears liquid crystal shutter glasses for an immersive experience \cite{Ohtani2021}. Recently, CompleXcope has been adapted for use with VR HMD \cite{OHNO2024}. Another approach, PlasmaVR, is being developed at Instituto Superior Técnico of the University of Lisbon in Portugal \cite{Trindade2024}. This system provides a 3D representation of plasma simulation data directly inside a VR HMD. Additionally, it offers interactive slicing and annotations through the 3D dataset, allowing users to view heatmaps of selected planes for a more detailed analysis of plasma behavior. Beyond VR, augmented reality has also been explored for plasma physics visualization using HoloLens \cite{Foss_2018,Mathur_2023}.

\section{Results} \label{sec3}
Below, we present various cases of high-intensity laser-matter interaction, categorized by topic, that have been simulated and visualized at the ELI Beamlines facility. The visualizations in the VR mode was tested using Oculus Rift S and HP Reverb G2 HDMs. The presented visualizations are  available online \cite{viz1,viz2,viz3,viz4,viz5,viz6,viz7,viz8,viz9,viz10,viz11}.
\subsection{Ion acceleration} \label{sec31}

 Laser-driven ion acceleration has remarkable potential applications, including medical treatment such as hadron therapy \cite{Bulanov2002,Bulanov2014,Tajima1997} and proton-boron capture therapy \cite{cirrone2018first,Istokskaia2023}, nuclear fusion \cite{Roth2001,Atzeni2002}, use in material sciences and nuclear physics research \cite{Nishiuchi2015}, as a neutron source \cite{Norreys1998,Horny2022}  and other areas \cite{borghesi2008,Daido2012,Macchi2013,Passoni2019}.  Non-destructive testing methods in cultural heritage investigations \cite{mirani2021integrated,barberio2019laser,passoni2019superintense} and environmental /forensic studies might also benefit from these advancements. Currently, laser driven ion acceleration is reaching energy levels of hundred of MeV per nucleon and beyond \cite{Higginson2018_100MeV,Rehwald2023,Ziegler2024}.
 
\subsubsection{Laser-Driven Proton Acceleration from Cryogenic Hydrogen Target} \label{sec311}

\begin{figure}[]
	\centering
	\isPreprints{}{
		\begin{adjustwidth}{-\extralength}{0cm}
			} 
		\includegraphics[width=1\linewidth]{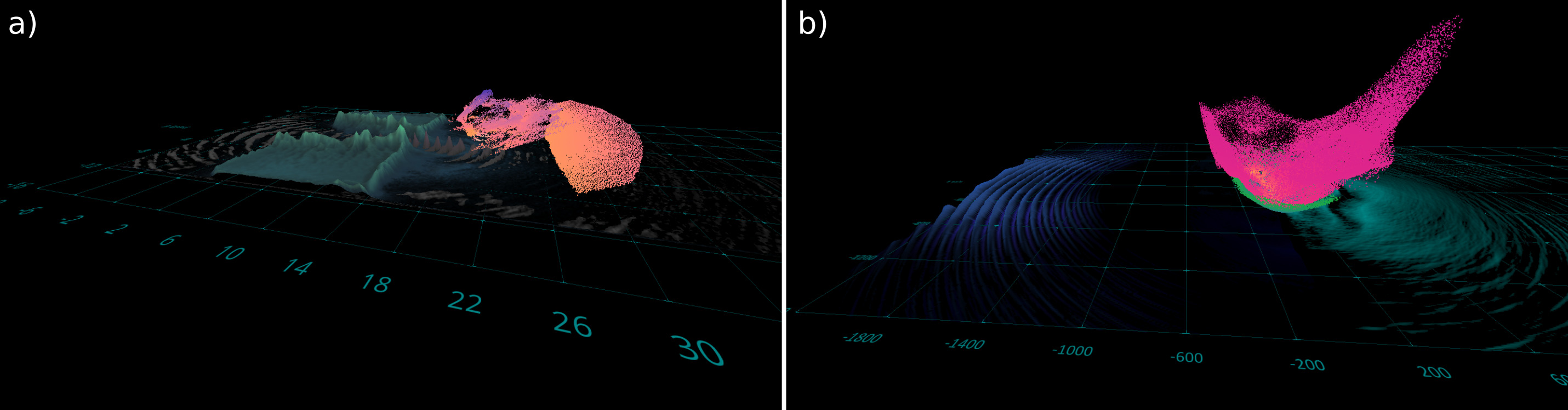}
		\isPreprints{}{
		\end{adjustwidth}
		} 
	\caption{Visualizations of laser interaction with a) cryogenic hydrogen target and b) plastic target for ion acceleration. The coloring is described in the text. The vertical height of the 2D layers represents the field amplitude, plasma density, and particle energy, respectively.}
	\label{fig_ion1}
\end{figure}

The visualization in Fig. \ref{fig_ion1}-a presents data from a 2D PIC simulation of a 9 PW laser pulse (with parameters relevant to the L4 laser at ELI Beamlines facility) interacting with a cryogenic hydrogen target. The laser pulse is depicted in grayscale, electron density in turquoise, and proton energy using a color scale ranging from white (zero energy) to purple ($\approx$400 MeV) and light blue (over 600 MeV). Only protons with energies exceeding 300 MeV at the end of the simulation are tracked, with their positions and energies visualized as dots.
Two distinct groups of protons, accelerated by different mechanisms, can be spatially distinguished: protons originating from the target interior, driven by Radiation Pressure Acceleration (RPA) \cite{Esirkepov2004}, and protons from the target’s rear side, accelerated via the Target Normal Sheath Acceleration (TNSA) mechanism \cite{Wilks2001,Snavelly2000}. In Refs. \cite{Psikal2018,Matys2023}, we demonstrated that under our conditions, RPA dominates over TNSA in both maximum energy and number of accelerated protons, a trend clearly observable in the time evolution of this VR visualization. Furthermore, in Ref. \cite{Psikal2018}, we analyzed the dependence of these mechanisms on laser intensity, polarization, and target material (using a plastic target), as well as the effects of introducing a short exponential preplasma on the target front side.

Cryogenic hydrogen targets have already been used in experiments involving lower-power and lower-intensity lasers \cite{Garcia2014,Margarone2016,Polz2019,Chagovets2022}, achieving proton energies up to 80 MeV \cite{Rehwald2023}. 

The visualization in Fig. \ref{fig_ion1}-a with its time evolution is available online at \cite{viz1}.

\subsubsection{Laser-Driven Ion Acceleration from Plastic Target}\label{sec312}

The visualization in Fig. \ref{fig_ion1}-b depicts the interaction of a high-intensity laser pulse with a micrometer-thick flat plastic target  obtained from a 2D PIC simulation using EPOCH code \cite{Arber2015}. The laser parameters correspond to those of the L3 HAPLS laser at ELI Beamlines facility \cite{sistrunk2017all,haefner2017high,L3HAPLS}.
The choice of flat mylar foil corresponds to the standard and widely-used experimental scenario.   The visualization clearly shows the acceleration of both protons (pink) and carbon ions (green), reaching maximum energies of 150 MeV/nucleon and 40 MeV/nucleon, respectively. Furthermore, different ion acceleration mechanisms can be distinguished: Hole-Boring RPA occurring at the front side of the target, and the dominant TNSA at the rear side. The presented visualization provides a better understanding of the energy, quantity, and directionality of all heavy particles, including the forward-moving accelerated proton and ion cloud/beam, as well as those moving backward. The latter is often overlooked in theoretical and numerical studies but is crucial for assessing potential damage to optical components in the interaction chamber located in the target-front vicinity (i.e., on the "laser side"). The follow-up 3D simulation of this configuration has been used to provide data to radiological protection studies \cite{bechet2016radiation} and Monte Carlo (MC) simulations of ELIMAIA user beamline \cite{margarone2018elimaia,ELIMAIA} in order to demonstrate the typical laser-ion acceleration scenario and beam transport through the ELIMED magnetic transport system and energy selector \cite{margarone2018elimaia,schillaci2015design}. Adjustments of laser-target parameters is ongoing research, which can enhance key features of ion beam , e.g., including maximum energy, particle number, spatial uniformity and homogeneity \cite{margarone2012laser,margarone2015laser}, as well as reduce ion beam divergence \cite{zakova2021improving}. 

Experiments using plastic targets have recently achieved comparable levels of proton acceleration, reaching up to 150 MeV \cite{Ziegler2024}.

The visualization in Fig. \ref{fig_ion1}-b with its time evolution is available online at \cite{viz2}.

\subsubsection{Collimated proton beam via double-layer target with modulated interface}\label{sec313}

\begin{figure}[]
	\centering
	\isPreprints{}{
		\begin{adjustwidth}{-\extralength}{0cm}
			} 
		\includegraphics[width=1\linewidth]{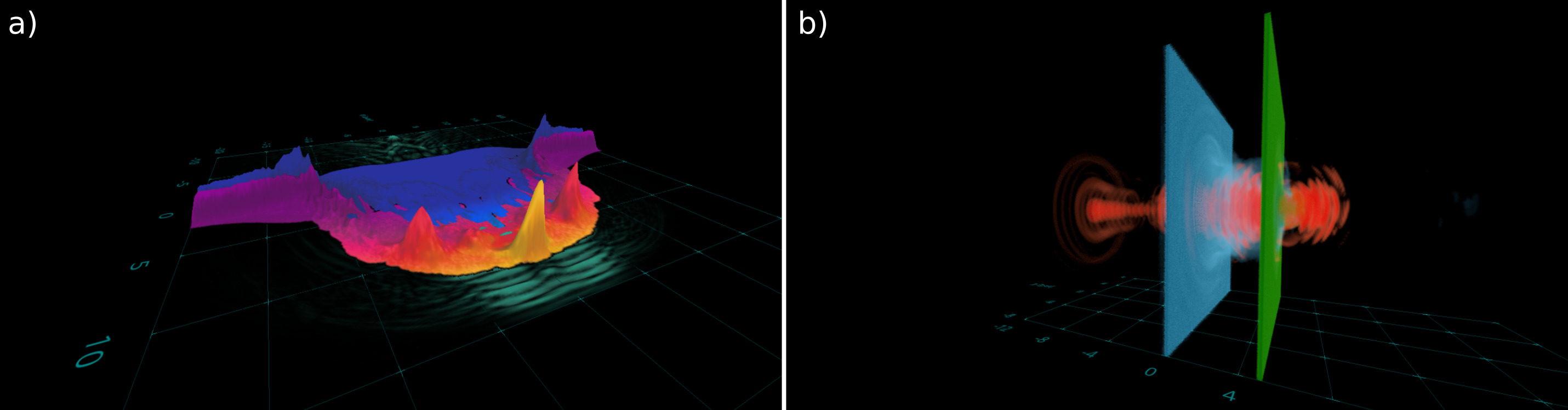}
		\isPreprints{}{
		\end{adjustwidth}
		} 
	\caption{Visualizations of laser interaction with a) double-layer deuterium-hydrogen target with corrugated interface b) plasma shutter and silver target for ion acceleration. The coloring is described in the text. The vertical height of the 2D layers in a) represents the field amplitude and ion densities, respectively.}
	\label{fig_ion2}
\end{figure} 

The visualization in Fig. \ref{fig_ion2}-a presents a 2D PIC simulation of a high-intensity steep front laser pulse (with parameters relevant to a 100 PW-class laser) interacting with a double-layer target featuring a corrugated interface. In Ref. \cite{Matys2020}, we demonstrated that the laser pulse (turquoise scale) drives a relativistic instability exhibiting Rayleigh-Taylor-like \cite{Rayleigh1882,Taylor1950} and Richtmyer-Meshkov-like \cite{Richtmyer1960,Meshkov1969} characteristics \cite{Wouchuk1996,Wouchuk1997,Nishihara2010,Mohseni2014,Matsuoka2017,Zhou2017,Zhou2019} at the interface between the deuterium layer (blue) and the proton layer (purple to yellow).

As the instability evolves, low-density plasma regions form alongside high-density ion bunches, with their positions determined by the initial interface corrugation. These bunches are then accelerated as compact structures by the laser radiation pressure, reaching energies of several GeV. Additionally, the laser pulse propagates through the low-density regions, generating an enfolding field around the central ion bunch. This field effectively confines the bunch, suppressing perpendicular expansion and leading to the formation of a collimated, quasi-monoenergetic proton beam of high energy (yellow). The presence of this beam is further confirmed in the proton energy spectra near the end of the simulation. 

Furthermore, in Ref. \cite{Matys2020}, we compared the performance of a corrugated double-layer target with other simulation configurations, including a double-layer target without interface corrugation and a single-layer pure hydrogen target. Our results demonstrated that the characteristic bunch structure in proton energy spectra and density distribution emerges only in cases where the interface is corrugated.

The visualization in Fig. \ref{fig_ion2}-a with its time evolution is available online at \cite{viz3}.

\subsubsection{Plasma shutter for heavy ion acceleration enhancement}\label{sec314}

The plasma shutter is typically a thin solid foil attached to the front surface of the target with a small gap between them \cite{Reed2009,Palaniyappan2012,Wei2017,Wei2025}. When a single foil is used without a secondary target, it is also referred to as a plasma aperture \cite{Gonzales2016electrons,Gonzalez2016proton,Jirka2021_nas}. As the laser pulse transmits through the opaque shutter, it develops a steep-rising front \cite{Vshivkov1998}, and its peak intensity increases locally at the cost of losing part of its energy. We demonstrated a local intensity enhancement by a factor of 7 \cite{Jirka2021_nas} and applied similar laser-shutter parameters in subsequent research involving a secondary target \cite{Matys2022}.

In Ref. \cite{Matys2022}, we investigated the application of plasma shutters for heavy ion acceleration driven by a linearly polarized high-intensity laser pulse, relevant to the L3 laser at ELI Beamlines facility. The visualization in Fig. \ref{fig_ion2}-b presents data from a 3D PIC simulation of a 1 PW laser (red) interacting with a silicon nitride plasma shutter (blue) and a silver target (green). The maximum energy of silver ions increases by 35\% (from 115 to 155 MeV/nucleon) when the shutter is included. Moreover, the application of the plasma shutter leads to ion focusing toward the laser axis in the plane perpendicular to the laser polarization, significantly reducing ion beam divergence. 

We also compared the effect for a circularly polarized pulse, which resulted in a 44\% increase in maximum silver ion energy. Furthermore, we observed the generation of a spiral laser pulse, which is further discussed in Section \ref{sec342}.

Additionally, we built a prototype of a double shutter and investigated it using a combination of 2D hydrodynamic and PIC simulations. Assuming a sub-nanosecond prepulse, the double shutter scenario increases the maximum energy by 260\% (from 64 to 167 MeV/nucleon) compared to the case without a shutter, where the prepulse remains unmitigated and the main target is pre-expanded.

The plasma shutters have been employed in lower-intensity experiments to generate a steep-rising front and increase laser intensity \cite{Palaniyappan2012,Wei2025}, and also to reduce the target pre-expansion by sub-ns prepulses, resulting in increased ion energies \cite{Reed2009,Wei2017}. For the acceleration of silver ions, energy exceeding 20 MeV/nucleon without a shutter was achieved \cite{Nishiuchi2020} using half the laser intensity and a target 25 times thicker than in our case. This is comparable to the 64 MeV/nucleon reached in our simulation with optimized parameters but without shutter, assuming a target pre-expanded by a 20 ps prepulse. Significant effort is currently being made to improve the prepulse characteristics of laser systems on the ns and ps levels \cite{Kiriyama2023}. Improvement of the fs prepulse and the generation of a steep-rising front in the main pulse will follow, where the plasma shutter is expected to play an important role, as demonstrated by our simulations \cite{Matys2022}.

The visualization in Fig. \ref{fig_ion2}-b with its time evolution is available online at \cite{viz4}.

\subsection{Electron acceleration} \label{sec32}
 Laser-driven electron acceleration is widely studied for its potential to create compact electron accelerators, with applications ranging from advanced radiation sources \cite{Corde2013, Albert2014} to investigations of strong-field quantum electrodynamics phenomena \cite{Gonoskov2022, Yu2024}.

\subsubsection{Electromagnetic-electron rings}\label{sec321}
High-power laser pulse propagation in low-density plasmas is central to various scientific challenges, including electron acceleration \cite{Tajima1979, Esarey2009}, the development of radiation sources \cite{Bulanov2013, Pirozhkov2012}, and nuclear fusion within the fast ignition concept \cite{Tabak1994}. For most of these applications, the laser pulse must propagate over extended distances while efficiently transferring energy into the plasma in a controlled way. In this context, much attention has been given to the evolution of the laser beam's radial profile in a fully ionized plasma, which can develop multifilament and, notably, ring-shaped transverse structures. Ref. \cite{Valenta2021} demonstrates that these electromagnetic rings can also generate high-energy ring-shaped electron beams.

\begin{figure}[]
	\centering
	\isPreprints{}{
		\begin{adjustwidth}{-\extralength}{0cm}
			} 
		\includegraphics[width=1\linewidth]{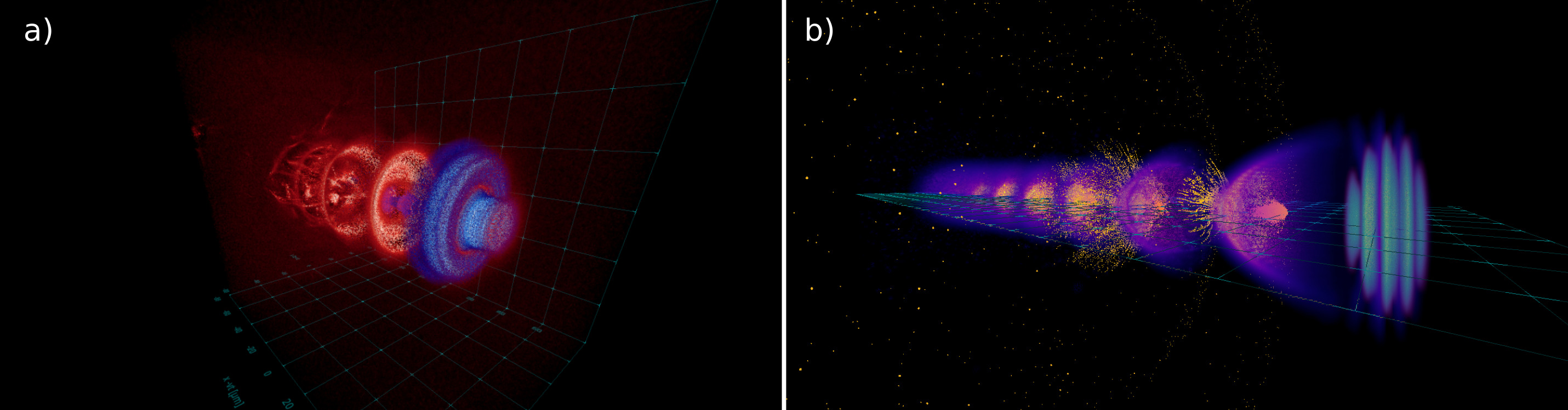}
		\isPreprints{}{
		\end{adjustwidth}
		} 
	\caption{Visualizations of a) Electromagnetic-electron rings, b) Nanoparticle-assisted Laser Wakefield Acceleration. The coloring is described in the text.}
	\label{fig_electron}
\end{figure} 

The visualization in Fig. \ref{fig_electron}-a presents a 3D particle-in-cell simulation demonstrating the formation of electromagnetic and electron rings. The laser intensity is displayed in blue, the electron density in red, and the accelerated electrons are represented by a violet-to-yellow colormap. Besides the applications mentioned earlier, understanding the mechanisms behind these ring structures is important for several reasons. First, electromagnetic rings can divert energy from the laser pulse, limiting the efficiency of high-intensity laser-matter interactions. Second, electron beams accelerated in their wake can damage surrounding equipment, such as capillaries used for guiding laser pulses, and generate unwanted radiation. Finally, identifying the conditions that lead to their formation can provide valuable diagnostics for studying high-intensity laser-matter interactions.

The formation of electromagnetic and electron ring structures in laser-plasma interactions has been experimentally observed, and several distinct mechanisms responsible for their origin have been identified \cite{Krushelnick1997,Kaganovich2008,Pollock2015,Yang2017,Behm2019,Salehi2021}.

The visualization in Fig. \ref{fig_electron}-a with its time evolution is available online at \cite{viz5}.

\subsubsection{Nanoparticle-Assisted Laser Wakefield Acceleration}\label{sec322}
To bring laser-accelerated electron beams to real-life applications, their stability and quality still need improvement. Producing such beams requires stable electron injection into the plasma wave (wakefield). Among various injection approaches, nanoparticle-triggered electron injection has emerged as a particularly promising mechanism, demonstrating consistent generation of high-quality electron beams in recent experiments \cite{Aniculaesei2023,Aniculaesei2019,Xu2022}. This method exploits the local field enhancement around the nanoparticle to trigger controlled injection at specific locations within the plasma, offering a potential path toward more reliable, reproducible, and even tunable electron beams suitable for practical applications \cite{Albert2016} in medical imaging, radiotherapy, and advanced materials science.

Many of these applications would benefit from high repetition rates, which conventional accelerators struggle to achieve. There are already several kHz laser systems available, offering energies in the range from several to a few tens of mJ. To reach intensities needed for electron acceleration with these systems, researchers must either use few-cycle pulses \cite{Salehi2021} or rely on small focal spots \cite{Lazzarini2024}. The latter was used to reach 50 MeV electrons, which is so far the highest electron energy achieved with kHz laser. Though with optimized electron injection even higher energies may be possible.

Figure \ref{fig_electron}-b presents results from a 3D PIC simulation of electron acceleration driven by a high-intensity laser pulse (displayed in green-blue), with electron injection facilitated by a 100 nm silica nanoparticle. The simulation was performed with a 7 fs, 10 mJ laser pulse focused to an intensity of $7\times10^{18}\ \mathrm{W/cm}^{2}$. The image captures the acceleration process 300 fs after the laser-nanoparticle interaction, when injection has already occurred. Electrons injected due to this interaction (depicted in pink-orange) form a well-defined bunch inside the wakefield structure, while electrons released during wave breaking (shown in yellow) appear dispersed throughout the surrounding plasma. Wave breaking refers to the process where plasma oscillations become nonlinear and break, similar to ocean waves on a shore, resulting in less controlled electron injection. This visual distinction clearly highlights the superior beam quality achieved through nanoparticle-triggered injection compared to self-injection mechanisms under these simulation conditions.

The visualization in Fig. \ref{fig_electron}-b with its time evolution is available online at \cite{viz6}.

\subsection{$\gamma$-flash generation and electron-positron pair production}\label{sec33}
The $\gamma$-flash radiation has exciting applications in various fields, such as materials science at extreme energy densities \cite{Eliasson2013}, $\gamma$-ray inspection and imagining \cite{Albert2016}, photonuclear
reactions \cite{Ledingham2000,Nedorezov2004,Kolenaty2022}, neutron sources \cite{Pomerantz2014}, photonuclear fission \cite{Cowan2000,Schwoerer2003}, radiotherapy \cite{Weeks1997}, shock-wave studies \cite{Antonelli2017}, quantum technologies\cite{Ujeniuc2024} and it can provide further understanding of the mechanisms of high-energy astrophysical processes \cite{Bulanov2015,Rees1992,Philippov2018,Aharonian2021,Esirkepov2012}. $\gamma$-photons are also used for subsequent electron-positron ($ e^- $--$ e^+$) pair generation \cite{Ehlotzky2009} (which can be further guided using orthogonal collision with another laser \cite{Maslarova2023,Martinez2023}). 

The possibility of positron source generation by exploiting the QED effects in ultra-intense laser-matter interaction has attracted a lot of attention \cite{Gonoskov2022}. This is important not only for fundamental research but also for applications, such as the non-invasive positron annihilation spectroscopy of materials \cite{Kim2024}.

\subsubsection{Collimated $\gamma$-flash emission along the target surface}\label{sec331}

\begin{figure}[H]
	\centering
	\isPreprints{}{
		\begin{adjustwidth}{-\extralength}{0cm}
			} 
		\includegraphics[width=1\linewidth]{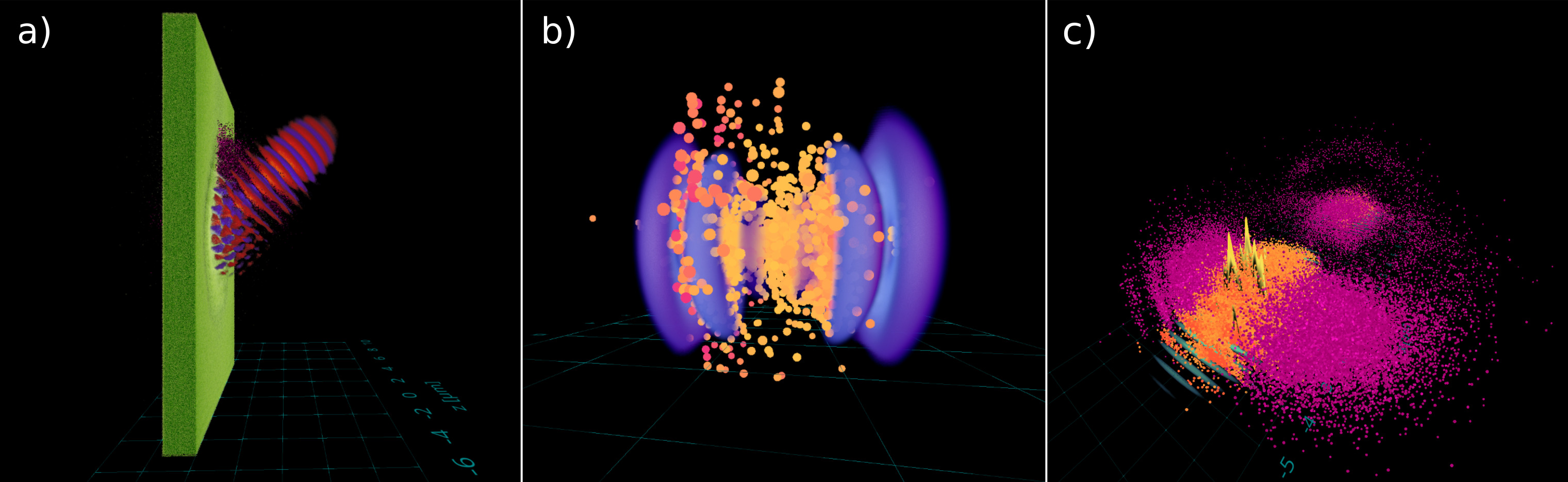}
		\isPreprints{}{
		\end{adjustwidth}
		} 
	\caption{Visualizations of a) Collimated $\gamma$-flash emission along the target surface, b) Electron-positron pair cascade in a laser-electron collision, c) Attosecond $\gamma$-ray flashes and electron-positron pairs in dyadic laser interaction with micro-wire. The coloring is described in the text.}
	\label{fig_gamma_pos}
\end{figure} 

A visualization of a 3D particle-in-cell simulation of a 14 PW laser (with red and blue representing $ \pm $ components of the laser electric field) interacting with an iron target (green) is shown in Fig. \ref{fig_gamma_pos}-a. In Ref. \cite{Matys2025}, we explored the effects of oblique laser incidence, where the incident and reflected parts of the laser pulse shape the electromagnetic field into a regular interference pattern. This field accelerates electrons to GeV energy levels, while simultaneously orienting their momentum in the direction parallel to the target surface. As a result, the electrons emit a collimated $\gamma$-photon beam (represented with the magenta scale) in the same direction. 

In Ref. \cite{Matys2025}, we also analyzed high-order harmonic generation in the vicinity of the target surface within our setup. Additionally, we examined the dependencies of $\gamma$-photon emission on various laser parameters, such as incident angle, polarization, power, and duration, as well as target properties, including thickness and preplasma conditions. These findings confirm the robustness of our proposed scheme.

The visualization in Fig. \ref{fig_gamma_pos}-a with its time evolution is available online at \cite{viz7}.
 
\subsubsection{Electron-positron pair cascade in a laser-electron collision}\label{sec332}
One widely accepted configuration for cascade pair production of electron-positron pairs is the interaction of seed electrons with multiple colliding laser pulses \cite{Blackburn2020, Gonoskov2022}.  As the electrons are trapped in the standing wave formed by two colliding laser beams, they emit photons by nonlinear Compton scattering. These photons being exposed to the strong laser fields are transformed into electron-positron pairs through the non-linear Breit-Wheeler process. The newly created particles again emit photons and the whole process is repeated.
However, to initiate the quantum electrodynamics (QED) cascade, the intensity of the order of $10^{24}~\mathrm{W/cm^2}$ is needed. Such a high laser intensity can only be achieved by tight focusing of the laser pulses. As a consequence of tight focusing, the strong ponderomotive force expels seed electrons from the interaction region and thus significantly suppresses or even prevents photon emission and pair production. Recently we have shown that the appropriate choice of laser pulse polarization can overcome this principal obstacle \cite{Jirka2024}. It is illustrated in Fig. \ref{fig_gamma_pos}-b showing the generation of QED cascade in a collision of two counter-propagating, tightly focused laser pulses with seed electrons initially located in the center of the simulation box. Radial polarization of the colliding laser pulses assures that the seed electrons are present in the interaction region even in the case of tight focusing. This feature allows the QED cascade to develop at 100$\times$ (80$\times$) lower laser power compared to the case of traditionally considered circular (linear) polarization.

The visualization in Fig. \ref{fig_gamma_pos}-b with its time evolution is available online at \cite{viz8}.
\subsubsection{Attosecond $\gamma$-ray flashes and electron-positron pairs in dyadic laser interaction with micro-wire}\label{sec333}

The 3D simulation of visualization shown in Fig. \ref{fig_gamma_pos}-c consists of three stages. At the initial stage, so called 'injection' stage, a radially polarized laser interacts with a micro-wire target, ejecting well defined microbunches of multi-MeV electrons. The laser used is assumed to be a TiSa laser of 25 PW power, focused by an f/4 parabola. The wire consists of lithium, with a diameter of 1.6 $\upmu$m and length of 3.2 $ \upmu $m. At the second stage, called 'boosting', those ejected electron bunches are further accelerated by the radially polarized laser, exceeding the GeV-energy range in high numbers and high collimation. A second 25 PW laser counter-propagates to the driving laser, with linear polarization and a tuneable f-number in the range f/0.8--f/20. When the linearly polarized laser interacts with the electron bunches, the so-called 'collision' stage occurs. There, for the high f-number cases (f/5--f/20 for the linearly polarized laser), an ultra-bright $\gamma$-flash is generated (purple dots in Fig. \ref{fig_gamma_pos}-c) due to the Compton scattering process. The resulting $\gamma$-flash consists of a series of attosecond $\gamma$-photon pulses, with a divergence of approximately 1 degree. By reducing the f-number, the interaction is dominated by Breight-Wheeler electron-position pair generation. The resulting  pairs (orange dots in Fig. \ref{fig_gamma_pos}-c) during the interaction time reach a density exceeding the solid density level, the most dense to ever be observed in the laboratory upon experimental realization of our proposed scheme.

The visualization in Fig. \ref{fig_gamma_pos}-c with its time evolution is available online at \cite{viz9}.

\subsection{Generation of attosecond and spiral pulse}\label{sec34}
Coherent electromagnetic attosecond pulses have impressive applications such as ultrafast coherent X-ray spectroscopy \cite{Ju2019}, non-linear XUV spectroscopy \cite{Bencivenga2015, Fidler2019}, and ultrafast X-ray protein nanocrystallography \cite{Chapman2011}.

The spiral pulses and spiral electron beams may play a crucial role in fast ignition \cite{Tabak1994} and ion acceleration \cite{Esirkepov2004}.

\subsubsection{Coherent attosecond pulse generation}\label{sec341}

Coherent electromagnetic attosecond pulses are typically facilitated through high-harmonic generation in partially-ionized noble gases, where phase-locked harmonics are produced in a train of attosecond pulses. While this method is currently the workhorse of attosecond science, it is limited in radiation intensity due to gas ionization limiting the maximum driving laser intensity. 

\begin{figure}[H]
	\centering
	\isPreprints{}{
		\begin{adjustwidth}{-\extralength}{0cm}
			} 
		\includegraphics[width=1\linewidth]{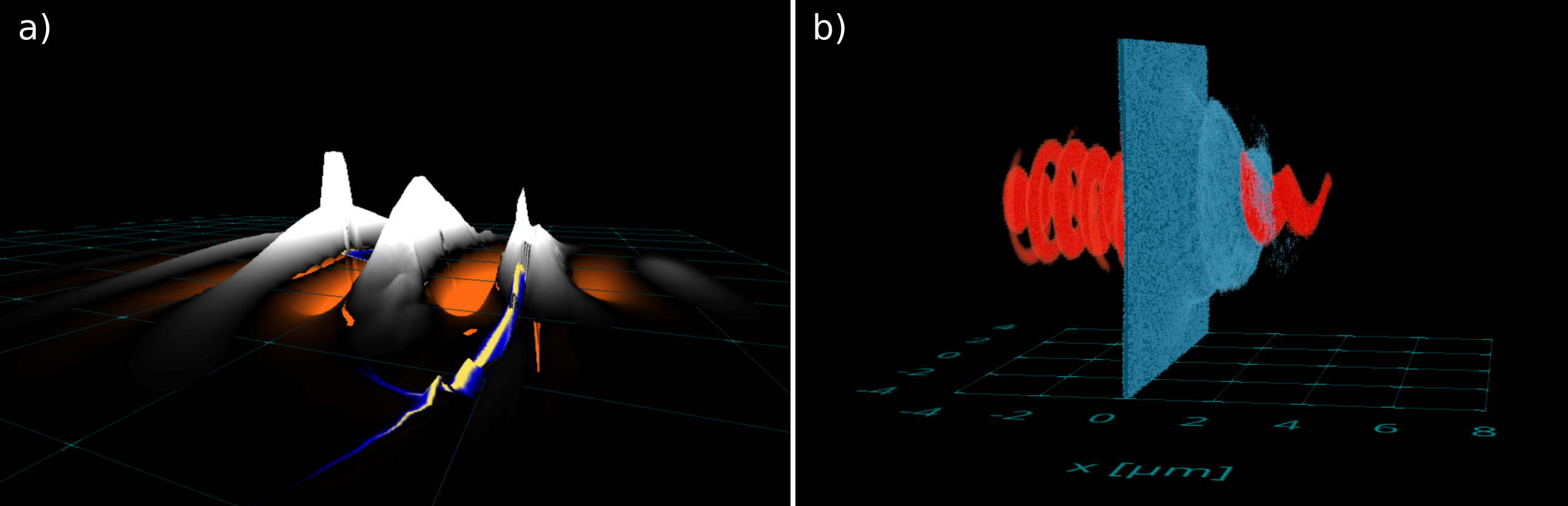}
		\isPreprints{}{
		\end{adjustwidth}
		} 
	\caption{Visualizations of a) Coherent attosecond pulse generation, b) Generations of spiral laser and electron beams. The coloring is described in the text. The vertical height of the 2D layers in a) represents the amplitude of positive and negative part of the field and ion density, respectively.}

	\label{fig_atto_spiral}
\end{figure}

High-harmonics obtained from relativistic oscillating mirrors \cite{Lamaifmmodecheckcelsevcfi2023, Quere2021, Edwards2020, Teubner2009, Bulanov1994} enable breaking through this limitation, as the plasma electrons conducting laser-driven oscillations are already ionized. Fig. \ref{fig_atto_spiral}-a shows a coherent attosecond pulse (white and orange colors depicts positive and negative parts of the magnetic field) produced from a relativistic plasma mirror (blue to yellow density scale) moving with constant velocity \cite{Einstein1905}, which can be driven either by charged particle beams \cite{Lamac2024} or intense laser pulses \cite{Bulanov2013, Valenta2020}.

The visualization in Fig. \ref{fig_atto_spiral}-a  with its time evolution is available online at \cite{viz10}.

\subsubsection{Generations of spiral laser and electron beams}\label{sec342}

In the visualization shown in Fig. \ref{fig_atto_spiral}-b, we examine hole-boring of a laser pulse through a silicon nitride foil, identical to the setup in Section \ref{sec314}, but this time using 3D PIC simulations with a circularly polarized laser pulse and no secondary target. As discussed further in Ref. \cite{Mima2024}, the circularly polarized laser (red) can be decomposed into a radially polarized wave and an azimuthally polarized wave within the hole. The radial polarization component generates spiral electromagnetic waves (red), which, through the Lorentz force, produce a spiral electron beam (blue). This electron beam, in turn, induces a strong quasi-static giga-gauss longitudinal magnetic field.

These spiral structures are inherently 3D phenomena and may play a crucial role in fast ignition \cite{Tabak1994} and RPA ion acceleration \cite{Esirkepov2004}. The structures exhibit rotational motion over time, as is seen by different time frames of the visualization, which are separated by one-quarter of the laser pulse period.

The visualization in Fig. \ref{fig_atto_spiral}-b with its time evolution is available online at \cite{viz11}.

\section{Discussion}\label{sec4}

In this work, we have presented the Virtual Beamline (VBL) application, an interactive web-based platform for the scientific visualization of high-intensity laser-matter simulations. Developed at ELI Beamlines facility, VBL integrates a custom-built WebGL engine with WebXR-based VR support, providing an immersive and interactive environment for data exploration. It supports two interaction modes: a non-VR mode and a VR mode. The non-VR mode operates entirely within a web browser and does not require any specialized hardware or software. The VR mode provides a fully immersive experience using a standard head-mounted display and its associated software. The framework efficiently processes and renders four main data types: point particles, 1D lines, 2D textures, and 3D volumes, using optimized transformation techniques to ensure smooth visualization.

By utilizing interactive 3D visualization, VBL overcomes the limitations of traditional 2D representations, offering a more comprehensive spatial understanding of complex plasma dynamics. The system supports multi-layer rendering and allows real-time manipulation of visualization parameters. Users can dynamically adjust time frames, switch between different data layers, switch colormaps and interactively explore data through moving, zooming and rotation. These interactive features enable researchers to analyze simulation results in greater detail, uncover hidden structures within the data, and better interpret complex plasma behavior. This functionality is particularly beneficial in both scientific analysis and educational outreach, making data exploration more engaging and accessible.

We have demonstrated how modern visualization techniques enhance the study of high-intensity laser-matter interactions, enabling the intuitive and dynamic representation of PIC simulations. The visualization framework has been applied to a variety of high-intensity laser-matter interaction scenarios, including ion acceleration, electron acceleration, gamma-flash generation, electron-positron pair production, attosecond and spiral pulse generation. These studies illustrate how VBL enables real-time exploration of complex simulation datasets, enhancing both research capabilities and public engagement with plasma physics.

Our repository of visualizations is hosted online and freely accessible via server \cite{vbl_home_page}, and it will continue to expand with new datasets from both simulations and experiments across different research fields. An example of external collaboration utilizing VBL for astrophysical phenomena visualization, based on simulations from Ref. \cite{Inchingolo2018}, was previously demonstrated in the earlier WebVR version of VBL \cite{Danielova2019}. We plan to continue creating new visualizations based on local simulation and experimental data at ELI ERIC, as well as from external sources upon agreement. As the database grows, this visualization repository will serve as a valuable resource for the user community of ELI ERIC \cite{eli_user} and beyond.


\vspace{6pt} 





\authorcontributions{M.M. wrote the bulk of the manuscript, coordinated the effort to revive the VBL project through a new version and implemented part of the updates, including improved online sharing capabilities and additional features, carried out the simulations and wrote texts in sections \ref{sec311}, \ref{sec313} ,\ref{sec314}, \ref{sec331} and \ref{sec342}, and designed the visualizations in sections \ref{sec322} and \ref{sec332} -- \ref{sec342}; J.P.T. implemented part of the updates in the new version of VBL, including WebXR support and additional features, and designed the visualization in section \ref{sec331}; M.K. coauthored the previous version of VBL, secured the documentation, scripts and knowhow transfer, and designed the visualizations in sections \ref{sec311}--\ref{sec314}; P. V. wrote the text, carried out the simulation and designed the visualization in section \ref{sec321}; M.G.Ž., A.Š., M.J., P.H., and M.L. wrote the texts and carried out the simulations in sections \ref{sec312}, \ref{sec322} and \ref{sec332} -- \ref{sec341}, respectively;  S.V.B. provided overall supervision. All authors have read and agreed to the published version of the manuscript.}

\funding{Portions of this research were carried out at the ELI Beamlines Facility, a European user facility operated by the Extreme Light Infrastructure ERIC.  
	The computational time was provided by the supercomputers Sunrise of ELI Beamlines facility and Karolina of IT4Innovations, supported by the Ministry of Education, Youth and Sports of the Czech Republic through the e-INFRA CZ (ID:90254). The support of Grant Agency of the Czech Technical University in Prague is appreciated, grants no. SGS22/185/OHK4/3T/14. Martina Greplová Žáková was supported by the Martina Roeselová Memorial Fellowship, provided by NF IOCB TEC-H.}

\dataavailability{The original visualizations presented in the study are openly available on https://vbl.eli-beams.eu/ }

\acknowledgments{We appreciate the work done on the previous WebVR version of VBL application by the former Virtual Beamline team, namely M.~Kecová, P.~Janečka, A.~Holý, K.~Petránek and J.~Grosz during their employment at ELI Beamlines. We appreciate the previous discussions with K.~Mima, T.~M.~Jeong, K.~Nishihara, J.~Pšikal, O.~Klimo and G.~Grittani regarding the simulations used for the visualizations, as well as the administration of the server hosting our webpage by P.~Wodecki and previously by J.~Majer.}

\conflictsofinterest{The authors declare no conflicts of interest.} 



\abbreviations{Abbreviations}{
The following abbreviations are used in this manuscript:\\

\noindent 
\begin{tabular}{@{}ll}
VBL & Virtual Beamline \\
VR & Virtual Reality \\
HMD & Head-Mounted Display \\
PIC & particle-in-cell \\
AI & artificial intelligence\\
RPA & Radiation Pressure Acceleration \\
TNSA & Target Normal Sheath Acceleration \\
QED & quantum electrodynamics \\
\end{tabular}
}


\appendixtitles{yes} 
\appendixstart
\appendix
\section[\appendixname~\thesection]{Controls of VBL application}\label{appendix}

Outside VR mode, navigation is controlled via the keyboard, with W, S, A, and D keys for translational movement and Up and Down Arrow keys or the mouse wheel for zooming. Rotation is controlled using two key pairs: the Q and E keys perform horizontal rotation, while the X and C keys control vertical rotation. Alternatively, the scene can be rotated by clicking and dragging with the left mouse button. The view can be reset to its default orientation by pressing the R key. Temporal navigation is managed using the Left and Right Arrow keys, which shift the visualization to the previous or next timeframe, respectively. The animation can be toggled (play/pause) using the Enter key. Individual visualization layers can be enabled or disabled using the number keys 0–9. The H and P keys toggle the display of the description and additional parameters. The Space key activates VR mode.

Inside VR mode, zooming is controlled via the A and B buttons on the VR controller, while translational movement is executed through a drag-and-pull motion with the right trigger. The Y button on the left controller pauses the visualization, and step-by-step time progression is managed via the left joystick. During playback, pushing the left joystick forward accelerates time, while pulling it backward rewinds it. Rotation in the horizontal plane is controlled by moving the right joystick left or right.

\begin{adjustwidth}{0\extralength}{0cm}

\reftitle{References}
\bibliography{references}
\end{adjustwidth}
\end{document}